# A note on the nature of de Broglie waves

M. R. Mahdavi

e-mail: mr_mahdavi@hotmail.com

In this note we will first argue that there is evidence in supporting the view that de Broglie waves have a gravitational origin. In view of the extreme weakness of the gravitational field, however, this seems to be an unlikely proposition. We consider the problem of the atom in the light of a proposed electron theory [physics/0212013] and show that in this theory an insignificant perturbation in the gravitational field might in effect be amplified by astonishingly large orders of magnitudes. On this basis we then propose a realistic model of an atom where the electron continually tunnels through the electrostatic field of the proton and the de Broglie waves take on the role of the missing electrostatic interactions. Although the discussion is a qualitative one, and as such may not be considered as being conclusive, never the less, we believe that on the evidence presented here the possibility of a gravitational origin for the de Broglie waves must be seriously considered.

Since the time Newton formulated his Mechanics, the mass of a body has, in one way or another, confronted physicist with insurmountable problems. Newton was puzzled by the equality of inertial and gravitational masses. Much later another problem arose when Lorentz introduced his electron theory [1], namely, the self-energy of a point electron came out as infinite. With the advent of quantum mechanics it was thought that the problem of the self- energy of the electron might at last be resolved. But, as early as 1929 it became clear that in quantum mechanics the self-energy of the electron would be infinite as well [2]. The problem of the infinities in quantum mechanics was never solved in a satisfactory manner; rather, it was sidestepped by a procedure that has come to be known as renormalization [3]. A basic assumption in the renormalization procedure is that what is important is the measured mass of the electron $m_e$, and not $M = (m_e + m_{el})$ the sum of its inertial and electromagnetic mass.

Renormalization finds its justification in removing the infinities and in predicting the very precise correction to the magnetic moment of the electron. However, one may ask what might be the underlying reasons for its assumptions, in particular the one made about the electron mass. We argue that in quantum mechanics it is the de Broglie waves that determine the trajectory along which an electron moves. Now, if de Broglie waves have a gravitational origin then it follows that an electron's own gravitational field determines the trajectory of its motion. Since the motion of a particle in a gravitational field is not dependent on its mass it then follows that the value of $M = (m_e + m_{el})$ is of no consequence. Thus, the inertial mass of an electron arises purely as a result of its being constrained to move along a guiding path, determined by its own gravitational field. An immediate consequence of this is that $m_e$ the inertial mass ought to be directly related to $m_g$ the gravitational charge of the electron. Hence, in this short argument one may discern an underlying reason for the equality of inertial and gravitational masses.

However, in view of the extreme weakness of the gravitational field, one cannot help but to have very serious doubts about the above proposition. Therefore, one has to show how in an atom for example, the insignificant gravitational fields may cause an electron to move in atomic "orbit". The discussion here will be based on the electron theory proposed previously [4] and a possibility that we considered in another paper [5]. In the following discussion we will adopt the terminology used in reference 4.

In [5] we showed that a non-spherical distribution of the electronic charge plus a sort of Brownian motion could result in situations where an electron and a nucleus would fail to see each other's electrostatic fields. In that event both particles can be considered as free ones as they do not undergo accelerations in each other's potential field. The Brownian motion, however, cannot be a purely random motion of an external origin as was to some extent suggested in that paper. For, there is an equal probability that such a random motion, which enables the two particles to tunnel through each other's electrostatic field, will also enable them to interact continuously with each other. Therefore, the random motion must be due to events having to do with the internal events of the quantum system in which the electron is situated.

In [4] we further developed the basic ideas presented in [5] and offered an electron model where the particle part of the electron, called the electrino, was periodically created and destroyed. However, in appearance the electrino moved in a circular motion. Obviously, the rapid oscillations of electrino would set up gravitational waves, and in the case of the electron at rest, a standing wave would appear around the electron. Now if we assume that quark models are constructed on the same principles as that of the electron, then at any instant an electron in an atom would be subjected to the insignificant waves emanating from the nucleus plus the remnants of waves created by itself at earlier times. These insignificant waves must now and in effect take on the role of the electrostatic interactions between the electron and the nucleus, which as required by the theory [5], must disappear from the scene.

Consider an electron moving towards a proton at an impact parameter d. Due to its spin and the annular shaped charge the electrino will interact intermittently with the proton and it will radiate energy. Along its trajectory it will reach a point A as shown in **Fig. 1**. At this point the orientation of the annular shaped charge in space is such that it is not subjected to the electrostatic field of the proton (in this figure we are looking at the annulus from an edge wise direction). In order that an atom may be formed we now require that, from hereon the electron will move in such a way that its electrino will always fail to "see" the electrostatic field of the proton. To see how this may come about we once again look at the interior of the electron.

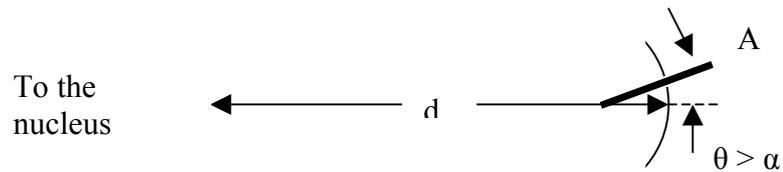

**Fig. 1**

In **Fig. 2** the trajectory of a $\psi^*$-field emanating from an electrino at point A is shown. This $\psi^*$-field is supposed to create a new electrino, after having looped around a black hole (see the addendum) at the centre of the circle, at point B at the instant that the first election's life-time comes to an end. The trajectory shown for the $\psi^*$-field, however, is for a smooth space. That is in travelling from A to the vicinity of the black hole near the centre of the circle, it will not pass any region where, there is an alteration in the space metric. Failing this, a slight change in the space metric at any point along the direction of the travel of the ψ*-field will bring about an imperceptible change in its direction of travel. Thus, consider that roughly the dimensions of the black hole is $\sim 10^{-33}$ cm. Therefore, an insignificant change in the direction of motion of the ψ*-field, say, of the order of much less than $10^{-20}$ radians relative to the direction of its undisturbed motion will bring it nearer to or further away from the edge of the black hole. Obviously the angle through which the ψ*-field loops around the black hole will critically depend on the distance of its nearest approach to the black hole horizon. The return part of the ψ*-field's trajectory will now be disproportionately altered. Thus a closer or more distant approach to the space warp horizon will result, for example, in new trajectories AOC or AOD, respectively. Thus a change in position of the order of $10^{-33}$ at the Planck scale may result in a change of the order $10^{-13}$ cm at the Compton scale. In this way the electron, at any instant that it might be on the verge of "seeing" the electrostatic field of the proton may dodge the electrostatic field by jumping to a new position in space.

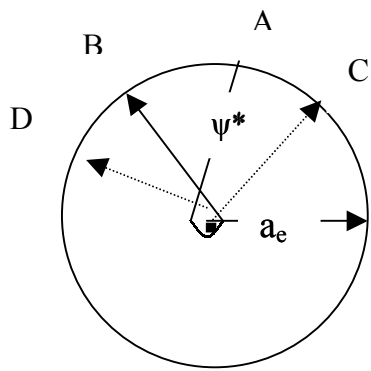

**Fig. 2**

In quantum mechanics in the case of an atom one writes down Schrödinger's equation and inserts the electrostatic potential in it. Having done this the solutions are obtained and they are interpreted as probability amplitudes. Something of a mysterious nature seems to have happened here. At the start we have electromagnetic interactions, but these seem to disappear once we have obtained a stationary state. Indeed, if we were to summarise the solutions of the Schrödinger equation in one simple sentence we would say that, "a stable stationary state of an atom is obtained if the value of the electric moment for the atom at all times is equal to zero". With spherically shaped charges this is impossible. However, it becomes possible if the elemental charges have annular forms and de Broglie waves are real physical waves, which in the effect that they produce can directly imitate the electrostatic interactions.

**Addenda:**

In [4] we emphasized that the black hole is not a black hole in the ordinary sense of the word and it only amounts to a temporary and local alteration in the metrical structure of space. We did not mention that such a black hole does not have long-range gravitational field associated with it, which might have confused the reader, mainly because this assertion, on the face of it, means that there two types of gravitational fields. This is not the case and we will come back to this point in a future paper.

In [4] we also hastily proposed that the black hole replenishment must be through the agency of the ψ*-field. This is obviously a mistake. The replenishment at any time can only be done by the remnants of space oscillations set up at earlier times and arriving in phase at the centre of the electron.

**References:**

[1]. Lorentz, H. A., *The theory of electrons,* Teubner (1916).
[2]. Heisenberg, W. and Pauli, W., *Zeits. Für Phys.*, **56** (1929) 1.
[3]. Dyson, F. J., *phys. Rev.* **75** (1949) 486 and 1736.
[4]. Mahdavi, M. R., physics/0211102 (2002).
[5]. Mahdavi, M. R., physics/0212013 (2002).